# 5G UE and Network Asset Administration Shells for the Integration of 5G and Industry 4.0 Systems

Jorge Gómez-Jerez, Jorge Cañete-Martín, M.Carmen Lucas-Estañ, Javier Gozalvez
*Uwicore laboratory, Universidad Miguel Hernandez de Elche*, Elche (Alicante), Spain
{jorge.gomezj, jcanete, m.lucas, j.gozalvez}@umh.es

*Abstract*—**5G is a fundamental technology for the full digitalization of smart manufacturing. The use of Asset Administration Shells (AAS) can facilitate the integration of 5G with Industry 4.0 systems and applications while minimizing the complexities associated with the 5G network management. This study presents the design and implementation of the first full 5G system AAS that is openly released to the community [1]. It includes a 5G UE (User Equipment) AAS and a 5G NW (network) AAS that have been designed following the 5G-ACIA guidelines as well as the Plattform Industrie 4.0 and 3GPP standards. The AASs have been defined to provide and expose the data and capabilities of 5G necessary to facilitate the integration of 5G with Industry 4.0 systems and applications.**

## I. Introduction

5G has been identified as a fundamental technology in the path towards the full digitalization of smart manufacturing and the foreseen evolution to flexible manufacturing. 5G can support the stringent latency, reliability and bandwidth communication requirements demanded by industrial applications, and solutions have been proposed to integrate 5G with Time-Sensitive Networks (TSN) deployed in factories. Furthermore, the possibility to deploy 5G Non-Public Networks (NPN) within factories increases the security and privacy guarantees to factory operators. However, the complexity to manage 5G networks can hinder their integration in industrial environments, and tools are needed to facilitate such integration while abstracting away from the complexity of the network management [2].

A widely adopted approach to facilitate interoperability and integration in industrial environments is the use of Asset Administration Shell (AAS) introduced by Plattform Industrie 4.0 [3]. An AAS is a standardized digital representation of an asset (structured into submodels), whereby assets in the context of Industry 4.0 can be machines, products, documents, software engineering tools or applications, among others. AASs enhance interoperability between different and heterogeneous industrial components and systems by providing structured data models and standardized interfaces for communication based on, for example, Open Platform Communications Unified Architecture (OPC UA) [4]. In this context, the definition of a 5G Asset Administration Shell (AAS) would facilitate the adoption and integration of 5G in industrial systems, and has been importantly promoted by industry-led organizations such as 5G-ACIA (5G Alliance for Connected Industries and Automation) [5]. It is interesting to highlight that a 5G AAS could be used to connect assets in the physical domain, between the physical and digital domains (in the context of digital twins), but also in the digital domain, e.g. through their integration into 3D manufacturing digital models.

5G-ACIA discusses in [5] the main aspects of 5G systems that should be considered for a structured description of the overall 5G network into an AAS. 5G-ACIA advocates for modeling the User Equipment (UE) separately from the rest of the components of a 5G network (NW). UEs may be provided by different suppliers and will be integrated into industrial devices of varying nature. Decoupling the modeling of UEs from the rest of a 5G system facilitates the interoperability and integration of 5G UEs into industrial devices and their corresponding AASs. In this context, 5G-ACIA proposes in [5] the design of a 5G UE AAS and a 5G NW AAS, and provides some guidelines about the most relevant information and parameters that should be included in the AASs. However, it does not provide a complete design for the 5G UE and NW AASs. The study in [6] proposes a Wireless Module submodel integrated into the AAS of an industrial machine, and discusses how the use of AASs facilitates the collaboration between automation and communication systems to enhance resilience. The Wireless Module includes the technical characteristics of the transmitter and receiver as well as measurements of the received signal level. The proposal in [6] was extended in [7] with a Quality of Service (QoS) submodel that contains parameters such as the response time and update time. However, the proposal still focuses on the device part (i.e. the UE) and does not include the network AAS. In addition, while it includes important wireless information, it is not 5G standard compliant. Another interesting contribution is the proposal in [8] where authors propose the definition of an Intent submodel within 5G UE and NW AASs for intent-driven 5G network management. An intent specifies a desired target state but does not establish how to achieve it. The purpose of this submodel is for industrial machines to specify their own network intents, and let the 5G NW AAS manage the network to achieve the specified intents. We should note that the AASs proposed in [6], [7] and [8] only include passive parts with properties and information that can only be read and written (and are not openly released). On the contrary, [9] defines a 5G submodel with active parts in the AAS that embed decision-making capabilities. In particular, [9] presents a submodel for the establishment or release of 5G connections (or PDU sessions), and introduces an operation to obtain subscription data for a specific UE from a 5G network. However, the proposal focuses only on this specific functionality of a 5G system.

Existing 5G AAS proposals focus on modeling specific aspects and functionalities of wireless systems and 5G, and the definition of a complete 5G AAS that includes all needs and requirements of Industry 4.0 applications is still necessary. In this context, this paper advances the state-of-the-art with the definition and implementation of a 5G system AAS that is 3GPP-standard compliant and follows the current AAS standards defined by Plattform Industrie 4.0 [3]. The proposed 5G system AAS follows the 5G-ACIA recommendations and guidelines in [5] and [10] to facilitate the integration of 5G with industrial systems and applications. To this aim, our 5G system AAS is made of a 5G UE AAS and a 5G NW AAS that provide access to, and expose, the main parameters and capabilities of 5G networks to support Industry 4.0 systems and applications. The proposed 5G UE and NW AASs include passive properties and define active operations for the interaction between 5G components as well as between 5G and industrial devices, systems, and applications. The AASs have been implemented using the AASX Package Explorer [11] and BaSyx [12] (in particular its SDK for Python), and their implementation and code are openly released with the paper in [1]. To the authors' knowledge, this is the first implementation of a 5G system AAS that is openly released to the community.

## II. Asset Administration Shell

An AAS is a standardized digital representation of an asset that provides relevant information about its features, capabilities, status, and operational data. In the context of Industry 4.0, assets encompass both physical and logical entities including machinery, documents, materials, and software. AASs also serve as virtual interfaces and facilitate the seamless integration and interoperability between different manufacturing components, systems, and applications. AASs represent a valuable tool to digitally manage and optimize industrial assets and manufacturing processes.

The information provided by an AAS is organized in digital models or submodels [1]. An AAS includes passive data that is used to represent the characteristics of the physical entity it represents. This data is represented as submodel elements and can be of different types, e.g. properties, lists, collections. The data is readable and/or modifiable via a message interface (e.g. an Application Programming Interface or API). The AAS may also incorporate an active part with decision-making capabilities that describes the operations that can be executed to realize certain technical functions or operations, e.g. reconfiguring a procedure, or modifying a service configuration. An AAS includes a message interface that supports interaction between AASs, and between AASs and software applications. The interface also enables controlled access to AAS data and operations. Plattform Industrie 4.0 proposes in [3] a classification of AASs based on types of interaction. Type 1 corresponds to passive AASs that exchange data in the form of a file using a standardized structuring of data, e.g. AASX, XML, JSON or AutomationML. Type 2 re-active AASs utilize standardized interfaces (REST or OPC UA) to interact using client/server or publisher/subscriber models. Type 3 pro-active AASs enable peer-to-peer interaction between AASs, which allows them to collaborate and modify their operations based on this interaction (e.g. in negotiation processes).

## III. 5G AAS

The integration of 5G into future factories requires a suitable description of the 5G system based on AAS principles for interoperability and interaction with AASs of industrial assets. This paper presents a complete AAS of a 5G system that has been defined and implemented following the 5G-ACIA guidelines in [5]. In this context, we define the AAS of a 5G system that is made of an AAS of the 5G UE and an AAS of the 5G network (NW). The 5G UE is the endpoint of a 5G link and a functional part of a 5G-capable industrial device. The 5G UE AAS should then be included as a submodel of the AAS of 5G-capable devices. The 5G NW AAS models the most relevant functions of the 5G Radio and Core networks. The 5G UE and NW AASs could be used to connect to the physical assets in the context of a digital twin, or they could be used as well to connect to a digital model of a production plant to integrate 5G connectivity. The interconnection between AASs and between AASs and assets in a 5G-enabled manufacturing environment is illustrated in Fig. 1.

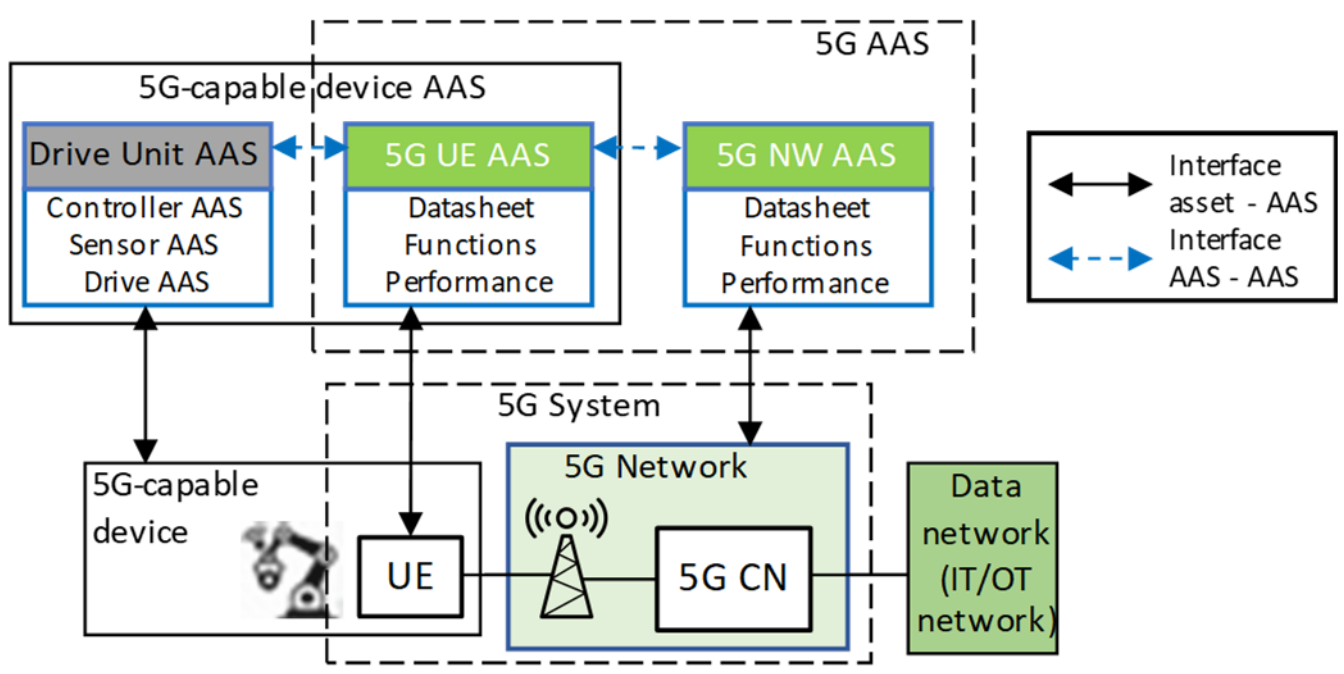

Fig. 1. Interconnection of digital and physical assets in a 5G-enabled manufacturing plant.

5G-ACIA discusses in [5] what data and properties should the AAS of a 5G system include based on 5G definitions within 3GPP. We build from these recommendations to present a complete structure for both the 5G UE and the 5G NW AASs. The definition of a 5G NW AAS can be a challenging task due to the complexity of 5G networks that are integrated by many physical (antennas, base stations, network servers, etc.) and logical elements (virtual RAN and CN network functions). We then propose a functional design of the 5G NW AAS where information is structured and grouped by functions or operations of a 5G network rather than by network nodes. This facilitates an easier evolution and scaling of the 5G NW AAS as the implementation can be tailored to the specific functionalities that are necessary or more relevant for each implementation of the 5G system AAS.

The proposed 5G UE and NW AASs model the main parameters, properties, and operations of 5G for its efficient integration with industrial or manufacturing processes. Such integration requires the capacity of industrial applications to access information about the 5G capabilities and performance [10]. To this aim, 5G defines analytics and exposure interfaces and services, including the Network Data Analytics Function (NWDAF), the Network Exposure Function (NEF), and the Service Enabler Architecture Layer (SEAL) [1]. Our AASs

[1] NWDAF collects data from various 5G core network and application functions, and derives statistics, e.g. related to network performance or network slicing [13]. These statistics can then be exposed to external applications using NEF [14] that provides various APIs for events

provide the capacity to retrieve data on demand, periodically or on an event-triggered basis following the exposure interfaces in 5G to facilitate the integration with industrial processes. In particular, both AASs will be able to interface with other AASs and industrial applications using OPC UA. The AASs utilize event subscriptions to expose or receive from other AASs data that may change dynamically. An event may be an on-demand request for data or periodic reports to monitor the evolution of the target data. It is also possible to trigger events when specific conditions occur, for example, when a parameter that is monitored exceeds a predefined threshold or a UE enters a certain area in the factory plant. To manage the event subscriptions, the submodels within our AASs contain a list of requested event subscriptions, and a list of notified events for each subscription. Maintaining a record of notified events allows tracing and analyzing the evolution of the 5G system, which is in line with the recommendations in [5] to maintain updated information about the AASs throughout their lifecycle.

The 5G UE and NW AASs presented in this paper correspond to type 2 or reactive AASs. The paper presents the complete structure and organization of the 5G UE and NW AASs in submodels including all the passive data along with the active operations that we propose should be included in the submodels. We should note that the proposed 5G UE and NW AASs have been designed also following the specifications of Plattform Industrie 4.0 for the definition of AASs in [3][16]. As a result, the 5G UE and NW AASs include, in addition to all 5G submodels that will be presented in the following sections, the following five default submodels for interoperability of the AASs [17]: Nameplate, Identification, Documentation, Service and TechnicalData submodels. Description and standardized templates for most of these submodels are available in the repository of Plattform Industrie 4.0 in [18], and are used in our 5G UE and NW AASs. Both AASs have been implemented using the AASX Package Explorer application [11] and BaSyx [12] (in particular its SDK for Python) which are open-source tools for creating and managing AASs. The implementation and code of our 5G UE and NW AASs are openly released with the paper in [1].

## IV. 5G NW AAS

Fig. 2 shows the proposed 5G NW AAS and its submodels that are next described. The default (Plattform Industrie 4.0) AAS submodels are not included in the figure but are available in the full implementation of our 5G system AAS released at [1].

### A. *NPN5GNWIdentity submodel*

This submodel includes information to identify the network, in particular, the Public Land Mobile Network (PLMN) ID or Non-Public Network (NPN) ID.

### B. *AssetServiceRegistry submodel*

This submodel contains information about the characteristics of the 5G network at planning and deployment phases following [5], including the Asset Service description, the identification of the Integrator Company and planning references. It also provides 5G coverage maps for the factory where the network is deployed, and information about the Service Level Agreements (SLA) between the 5G network operator/provider and the factory operator. The SLA includes the metrics by which the level of service is measured, and the expected performance per service type.

### C. *Network5GDataSheet submodel*

This submodel includes information that characterizes the technical capabilities of the deployed 5G network, including the supported 3GPP release, the supported and used spectrum bands, the maximum data rate achievable in downlink (DL) and uplink (UL), and the supported network protocols (IPv4, IPv6, etc). The submodel also provides information about the network topology, and includes a list of all RAN and CN nodes that comprise the deployed 5G network. For each RAN and CN node, the submodel indicates the type of node (gNB, UPF, PCF, SMF, etc.), its IP address, its connections with other nodes, its location, the virtual machine hosting the node (if this is the case), and the computing and memory resources allocated at the virtual machine. In the case of RAN nodes, the submodel also includes information about the maximum transmission power, the spectrum band, and the receiver sensitivity. The submodel also includes information about the links between nodes as well as the links' capacities.

### D. *VirtualNetworks submodel*

5G allows the implementation of virtual or logical networks to support different QoS (Quality of Service) levels. One option is network slicing, which exploits the virtualization and softwarization of networks to create different logical networks or slices over a common network infrastructure. Network slicing is considered one of the main 5G technologies to support industrial services with diverse and stringent requirements using a single 5G network. Network slices (NS) are configured to support specific QoS profiles. A 5G QoS profile defines a set of QoS parameters and characteristics that determines the treatment of the data traffic in the 5G network. 5G can also implement Virtual Local Area Networks (VLAN) to manage separately different traffic flows. VLANs operate at the data link layer and are logical Ethernet networks that are created over the same physical infrastructure by assigning ports on network switches to specific VLANs. In a VLAN, traffic is separated by including the VLAN ID in each frame.

The *VirtualNetworks* submodel contains the list of NSs and VLANs created. For each NS, the submodel indicates the Single Network Slice Selection Assistance Information (S-NSSAI) [19] as well as the configurable attributes and their values. The S-NSSAI identifies the NS and the Slice/Service type (SST). The SST defines the expected NS behaviour in terms of features and services. 5G defines standardized SSTs for enhanced Mobile Broadband (eMBB), ultra-reliable low latency communications (URLLC), massive IoT (MIoT), V2X and High-Performance Machine-Type Communications (HMTC) services in [19]. The attributes that characterize a NS are defined in [20], and include, among others, availability, area of service, isolation level, UE density, and UL throughput per NS (see [20] for a complete list). [20] also establishes the minimum attribute values for the standardized SSTs that can be configured through our *VirtualNetworks* submodel. The submodel also identifies the computing resources of the virtual machines where the NS is executed. For VLANs, the monitoring or analytics exposure. 5G also simplifies access to information commonly used to develop vertical applications (e.g. group, location, identity and network resource management) through the SEAL enablers [15].

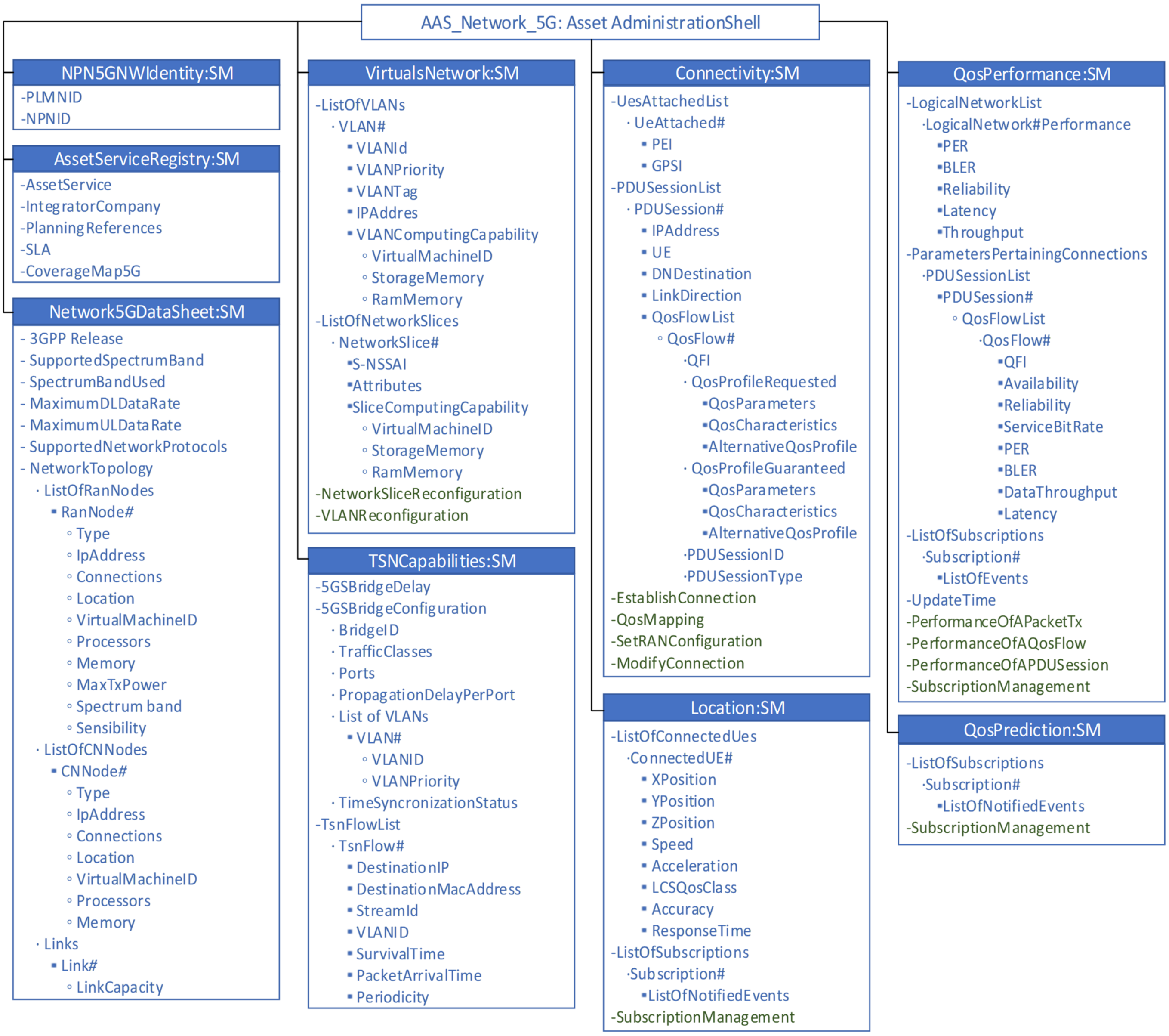

Fig. 2. 5G NW AAS and submodels.

submodel indicates the IP address, VLAN ID, priority, VLAN tag (identifies the VLAN in a data frame), and allocated computing resources to execute the VLAN.

The *VirtualNetworks* submodel has an active part to implement the *NetworkSliceReconfiguration* and *VLANReconfiguration* operations. These operations should emulate the reconfiguration process of a VLAN or network slice respectively, based on, for example, new network requirements. For example, a NS reconfiguration might occur if a Reinforcement Learning process is implemented to monitor the performance of a 5G NS and dynamically adapts the NS's attributes if the expected performance is not achieved. The operation will return the new configured parameters of the slice if the 5G network accepts the reconfiguration request.

### E. *Connectivity submodel*

The *Connectivity* submodel provides information about the UEs that are currently connected to the 5G network. In particular, it includes a list of attached UEs identified by their Permanent Equipment Identifier (PEI) and Generic Public Subscription Identifier (GPSI). The submodel also includes a list of the established PDU (Packet Data Unit) sessions. A PDU session is a logical connection within the 5G system between a UE and the user plane function (UPF) that provides access to a Data Network (DN). The submodel includes per PDU session the list of established QoS Flows. Each QoS flow is identified by a unique identifier called QoS Flow ID (QFI), and by a QoS profile that includes the QoS parameters that characterize the packet flow as defined in [19]. The QoS parameters include the 5G QoS Identifier (5QI), Allocation and Retention Priority (ARP), Reflective QoS Attribute (RQA), Guaranteed and Maximum Flow Bit Rate (GFBR and MFBR, respectively), Notification control, Maximum Packet Loss Rate, and Aggregated Bit Rate. The 5QI is a scalar value that identifies a series of QoS characteristics that should be guaranteed for a QoS Flow with the corresponding QoS profile. These characteristics include, among others, the resource type, priority level, packet delay budget or PDB (i.e. the latency requirement), packet error rate (PER), averaging window and maximum data burst volume. The submodel includes the QoS profile requested in the PDU session establishment procedure, and the QoS profile guaranteed by the network after the QoS mapping process. The requested and guaranteed QoS profiles might not match if the network does not have sufficient resources. In this case, the industrial applications might adapt (if possible) to the guaranteed QoS profile or an alternative QoS profile might be used. This alternative QoS profile can be provided at the PDU session establishment. It represents alternative values for the PDB, PER and GFBR to which the application can adapt in case the requested QoS profile is not fulfilled [19].

The *Connectivity* submodel lists four possible operations. The *EstablishConnection* operation should emulate the establishment procedure of new PDU sessions [21]. It can be called, for example, when a new UE is attached to the network and needs to establish a new PDU session. The *QosMapping* operation should implement the QoS mapping process carried out in 5G (defined in [21]) when a new QoS profile is requested. The process considers the current state of the network to derive the guaranteed QoS profile. The *QosMapping* operation could also be executed to (re-) negotiate the specific QoS parameters of a new or already established connection if the UE QoS requirements change. The *Connectivity* submodel also includes the possibility to implement a *SetRANConfiguration* operation. This operation establishes the value of different RAN parameters (at both the 5G NW and 5G UE AASs) that are important to support specific QoS profiles, including among others, the use of a specific or a set of Modulation Coding Schemes (MCSs), the maximum number of retransmissions or the 5G numerology. The *SetRANConfiguration* operation should be executed before starting the 5G transmissions in a PDU session. It could also be executed to evaluate what performance could be achieved with a specific configuration of the RAN parameters. Finally, the *Connectivity* submodel includes the *ModifyConnection* operation to emulate the PDU session modification procedure in 5G [14]. This procedure is executed to modify the configuration, in particular the QoS profile, of an already established connection or PDU session. The procedure can be initiated by the network or following a request from a UE.

### F. *QosPerformance submodel*

The *QosPerformance* submodel provides information about the performance experienced for a specific packet transmission, a QoS Flow or a PDU session. The performance can be expressed in terms of [22]: service availability (percentage of time that the service is delivered according with the guaranteed QoS profile), reliability (percentage of packets successfully delivered within the latency requirement or PDB), Packet Error Rate (PER), service bit rate, Block Error Rate (BLER), data throughput and latency. The submodel also indicates the update time that defines the periodicity with which the performance is calculated. The *QosPerformance* submodel also provides information about the performance achieved in each virtual or logical network (network slice or VLAN), in particular, the experienced PER, BLER, reliability, latency and throughput. This information is important to identify any malfunction or incorrect network configuration. The performance of the virtual or logical networks is also computed periodically based on the update time parameter configured in the submodel.

The submodel provides a list of operations to estimate the performance experienced for a specific packet transmission (*PerformanceOfAPacketTx*), a QoS Flow (*PerformanceOfAQoSFlow*) or a PDU session (*PerformanceOfAPDUSession*). With these operations, it is also possible to provide information about the performance experienced by one or a group of UEs during a given period. Finally, the *QosPerformance* submodel contains a list of subscriptions and a list of notified events. An external application or AAS can subscribe to receive information periodically, on demand, or on an event-basis about the QoS performance of 5G connections or PDU sessions, leveraging the 5G exposure capabilities introduced with NWADF, NEF and SEAL. In addition, the submodel includes the *SubscriptionManagement* operation to manage subscriptions requested by UEs or vertical applications.

### G. *TSNCapabilities submodel*

5G has been designed with the necessary enablers and functionalities to integrate with Time Sensitive Networking (TSN) networks and jointly support Time Sensitive Communications (TSC) [19]. In integrated 5G and TSN networks, the 5G network acts as a TSN bridge, also referred to as the 5GS Logical Bridge [19]. In this context, it is important that the 5G NW AAS includes a *TSNCapabilities* submodel like any other AAS of a TSN bridge to identify the 5G network as a TSN-capable device. This submodel includes information specific for a 5GS bridge like the 5GS Bridge delay and the propagation delay per port, as well as configuration parameters including the IP, ports, VLAN-ID and VLAN priority. The submodel also includes a list of the active streams that correspond to different TSN flows. For each TSN flow, the model provides information about the parameters included in the TSCAI or TSC Assistance Information (survival time, packet arrival time and periodicity). We should note that the TSCAI parameters are provided by the TSN network so that the 5G system can adequately support the TSN traffic.

### H. *Location submodel*

The 5G New Radio (5G NR) can significantly improve the positioning accuracy compared to previous mobile generations, which opens the door for the use of 5G positioning in vertical applications (including Industry 4.0). The 5G NW AAS includes a *Location* submodel to register the position of all UEs connected to the network. Each UE is identified using its PEI, and the submodel provides the position using Cartesian coordinates, along with speed and acceleration information. The submodel also includes the Location Service QoS Class (LCS QoS Class) [23] that indicates the positioning accuracy and response time to the positioning request tailored to the specific requirements of each application/UE [23]. The submodel also includes a list with subscriptions to different location data and events that can be provided on-demand, periodically or an event-triggered basis, along with a list of the notified events. For example, a UE can subscribe to periodic reports of its location, or an industrial application can subscribe to notifications when one or a group of UEs enter a given area. The submodel also includes the possibility to implement a *SubscriptionManagement* operation to manage the subscriptions to location events requested by UEs and vertical applications.

### I. *QosPrediction submodel*

With NWDAF, a 5G network can provide not only analytics but also the capacity to predict the QoS of 5G connections. Such predictions will be crucial for a proactive management of the 5G network that can anticipate the QoS performance and/or connectivity needs of the production processes it should support, e.g. based on the mobility of UEs, their traffic demand or the network status. Our current implementation of the 5G NW AAS corresponds to a type 2 or reactive AAS. However, our 5G NW AAS includes a *QosPrediction* submodel to facilitate its future evolution to a type 3 or pro-active AAS.

The submodel includes a list of subscriptions requesting either network QoS predictions or QoS predictions of specific UE connections. For each subscription, the submodel includes the QoS parameters to predict, the period of time for the prediction, and if applicable, information about the area or path of interest where the predictions are needed. Industrial applications can subscribe to receive QoS prediction events on-demand, periodically or an event-triggered basis. The submodel also includes a list of the notified QoS prediction events. The submodel includes a *SubscriptionManagement* operation so that a UE or external application can request a subscription to a QoS prediction event, specifying whether the subscription is for an on-demand, periodic or event-triggered notification.

## V. 5G UE AAS

The 5G UE AAS enables the effective management and monitoring of a 5G UE that is part of a 5G-capable industrial device. By providing a comprehensive digital representation of the UE, the 5G UE AAS facilitates the seamless integration of 5G into production networks. The 5G UE AAS is defined following a functional approach like in the case of the 5G NW AAS. Fig. 3 shows the proposed 5G UE AAS with its submodels that are presented in the next sections.

### A. *UE5GIdentification submodel*

This submodel contains different UE identifiers, including the IMSI (International Mobile Subscriber Identity), the PEI (Permanent Equipment Identifier), and the GPSI (Generic Public Subscription Identifier). The IMSI is a unique (and not public) identification code for each UE in a mobile network. The PEI is an identifier of the terminal equipment used by the UE. The GPSI is a public identifier used both inside and outside of the 3GPP system, for example, our public mobile phone numbers or mobile subscriber ISDN number (MSISDN). The submodel also includes the UE's PIN (Personal Identification Number) code, the ICCID (Integrated Circuit Card ID) that is a unique global serial number of the SIM (Subscriber Identity Module) card used by the UE, and the Service Provider Name (SPN) that identifies the serving mobile operator. Finally, the submodel includes the authentication certificate (and its status) used to authenticate a UE in a 5G network, as well as the IP and MAC addresses of the UE device.

### B. *UE5GDatasheet submodel*

This submodel includes technical characteristics of the 5G UE, including the operating bands and duplex mode it supports and parameters of the communications channel such as the maximum transmission bandwidth that the UE can support. The submodel also includes radio transmission and reception characteristics of the UE [24], including the maximum output power, output power dynamics, output RF spectrum emissions, reference sensitivity power level, adjacent channel selectivity, and number of transmission and reception antennas and layers.

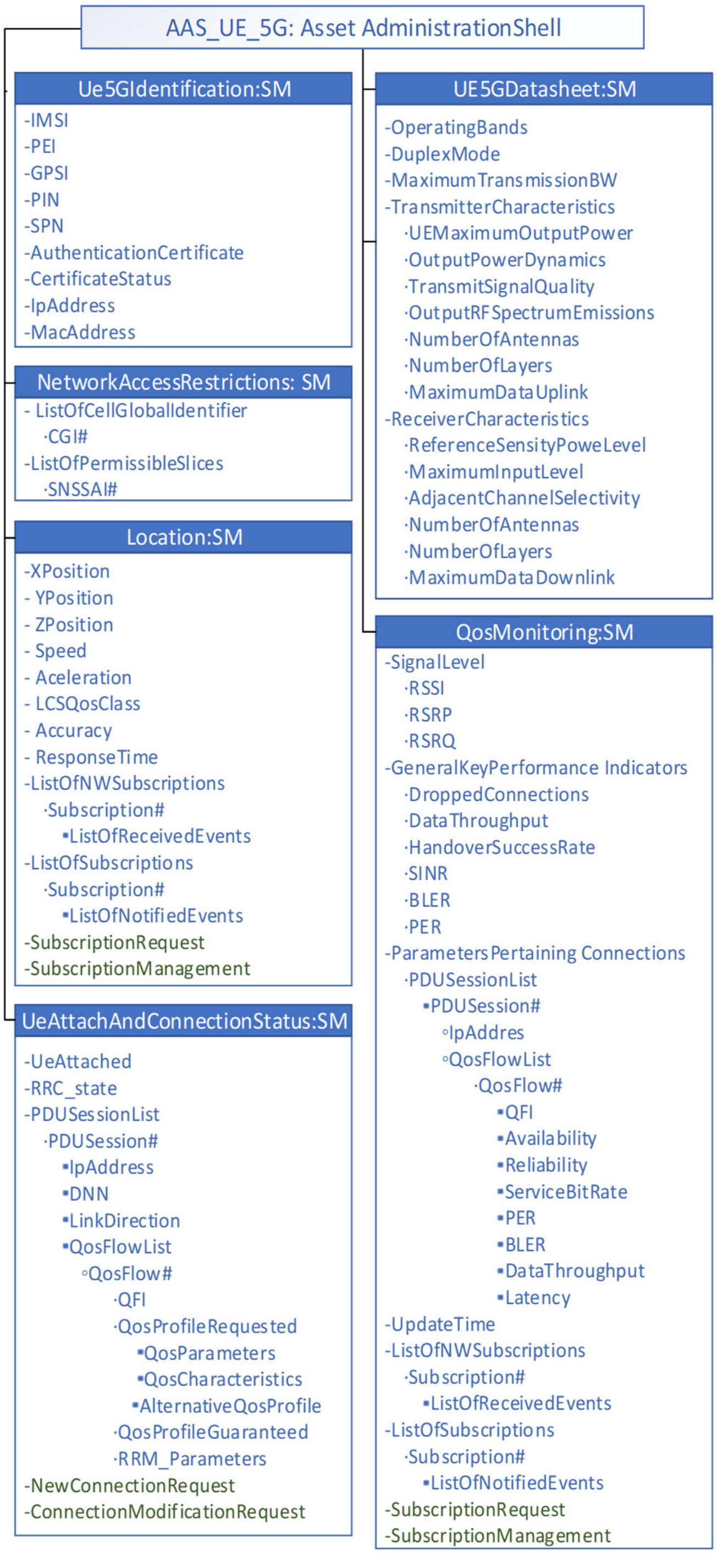

Fig. 3. 5G UE AAS and submodels.

### C. *NetworkAccessRestrictions submodel*

This submodel contains information related to the physical and logical access restrictions of the UE to the 5G network. The submodel stores the list of Cell Global Identifiers (CGIs) that identifies the 5G cells that the UE can connect to. It also has a list of the network slices (identified with their S-NSSAIs) that the UE can connect to. This list is determined by the 5G network together with the guaranteed QoS profile [25]. It is important that the slices to which the UE can connect are capable of supporting the guaranteed QoS profiles for the UE.

### D. *UEAttachAndConnectionStatus submodel*

This submodel includes the UE connection status. In particular, it indicates if the UE is attached or not to the network, and the RRC (Radio Resource Control) state that indicates whether the UE is active, inactive, or idle. The submodel also includes a list of the active PDU sessions and QoS flows of the UE. For each PDU session and QoS flow, it also includes information about the QoS profiles requested and guaranteed by the 5G network. The submodel also includes the value of different Radio Resource Management (RRM) parameters necessary to satisfy each guaranteed QoS profile. These parameters are the MCS and Channel Quality Indicator (CQI) tables, the target BLER, the scheduling type and policy, the maximum number of retransmissions for Hybrid Automatic Repeat Request (HARQ), the number of repetitions ($k$), and the maximum transmission power for power control. We should note that the information related to the PDU sessions established by the UE is also contained for all connected UEs in the *Connectivity* submodel of the 5G NW AAS. However, we believe that it must be also accessible through the 5G UE AAS so that industrial applications implemented in 5G-capable devices can access this data.

The submodel includes the possibility to implement two operations. The *NewConnectionRequest* operation can be used by a UE to request the establishment of a new PDU session. The request includes the QoS profile requested by the UE. The submodel also includes the option to implement a *ConnectionModificationRequest* operation. The UE can use this operation to request modifying an established connection or PDU session, particularly its QoS profile guaranteed.

### E. *QosMonitoring submodel*

This submodel contains information about the performance experienced by the UE. It includes information about the signal quality received by the UE, including the RSSI (Received Signal Strength Indicator), RSRP (Reference Signal Received Power), and the Reference Signal Received Quality (RSRQ). The submodel contains information related to the performance experienced over all the PDU sessions established [10]. In particular, it contains the average data throughput, the percentage of dropped connections, the handover success rate, the Signal Interference Noise Ratio (SINR), and the average BLER and PER. The submodel also contains information about the performance experienced per PDU session and QoS flow (service availability, reliability, PER, service bit rate, BLER, data throughput and latency). The submodel also indicates the update time. We should note that the performance information for all PDU sessions and QoS flows per UE is also accessible through the *QosPerformance* submodel of the 5G NW AAS.

The submodel contains a list of requested subscriptions to QoS performance events requested to the 5G NW AAS, and a list of received events notified by the 5G NW AAS. The submodel includes a *SubscriptionRequest* operation to request to the 5G NW AAS a new subscription to QoS performance events on-demand, periodically or an event-triggered basis. This allows the possibility to continuously monitor if the QoS requirements of the UE are met (for instance, if the experienced latency exceeds a threshold or if a minimum service bit rate is not guaranteed), and to manage and optimize the connections (and if necessary, the network configuration) accordingly. The submodel also contains a list of subscriptions to QoS performance events requested to the 5G UE AAS by industrial applications implemented in the industrial device-side, and a list of notified events for each subscription. The industrial applications might subscribe to QoS performance events in the UE to, for example, adapt the operation mode of the application to changes in the QoS experienced. The *QoSMonitoring* submodel also includes the *SubscriptionManagement* operation to manage the subscription requested by industrial applications.

### F. *Location submodel*

It is important that the location information generated by the network is transmitted to the UEs and is made accessible to 5G-capable industrial devices. In this context, we define a *Location* submodel in the 5G UE AAS that contains information about the current UE location and the required location service quality (indicated by the LCS QoS Class, accuracy and response time [23]). It also provides a list of subscriptions to location information provided by the 5G network, a list of received events from the 5G NW AAS, and a *SubscriptionRequest* operation to request a new subscription to a location event to the 5G NW AAS (for example, changes in the device connection or periodic location reports). In addition, the *Location* submodel also contains a list of subscriptions to location events requested by industrial applications implemented in the device, and a list of notified events. This submodel also includes a *SubscriptionManagement* operation to manage the subscriptions to location events requested by the industrial applications. These subscriptions allow industrial applications to consider the location on their operation. Like for other submodels at the UE, we should note that the 5G NW AAS also contains the location information for all the UEs connected to the network.

## VI. 5G AAS Application Example

This section illustrates an example of the potential use of the defined 5G UE and NW AASs to support industrial processes. We consider an evolution of a use case proposed by 5G-ACIA in [10] that deals with QoS management in a 5G network supporting a production process. The use case is illustrated in Fig. 4. It considers a factory shopfloor where there are two 5G-capable industrial robots that have active connections with the 5G NW using their 5G UE AASs (UE1 and UE2). Each of these active connections has a guaranteed QoS profile. A new production process starts and requires high-resolution image inspection. A 5G inspection camera is activated to transmit the images to an image analysis unit on the cloud for automated inspection. The UE associated to the 5G camera (UE3) requests a new connection (or PDU session) with a high-bandwidth QoS profile (*NewConnectionRequest* operation in the 5G UE3 AAS in Fig. 4). Upon reception of the new connection request, the 5G NW AAS initiates the *QosMapping* operation to determine the QoS profile that can be guaranteed for the new connection based on the aggregated information about the current state of the network, the QoS requested (including its priority), and the QoS profiles of the active connections. The 5G NW AAS informs the 5G UE3 AAS of the guaranteed QoS profile, and the 5G NW AAS establishes the new connection for UE3 (*EstablishConnection* operation). The activation of UE3 changes the network status, and the 5G NW AAS sets new RAN parameters for UE2 (using the *SetRANConfiguration* operation in the 5G UE AAS) to maintain its guaranteed QoS profile, e.g. it increases the number of retransmissions. In addition, the 5G NW AAS needs to change the QoS profile for the UE1 using the *ModifyConnection* operation in 5G NW AAS. After all changes, the 5G NW AAS estimates the performance of all three active connections to assess whether the QoS requirements are met (*PerformanceOfAPDUSession* operation). The QoS performance experienced is updated in the 5G UE AASs (*QosPerformance* submodel) and 5G NW AAS (*QosMonitoring* submodel), and is made accessible to industrial applications or devices through the corresponding subscriptions and events.

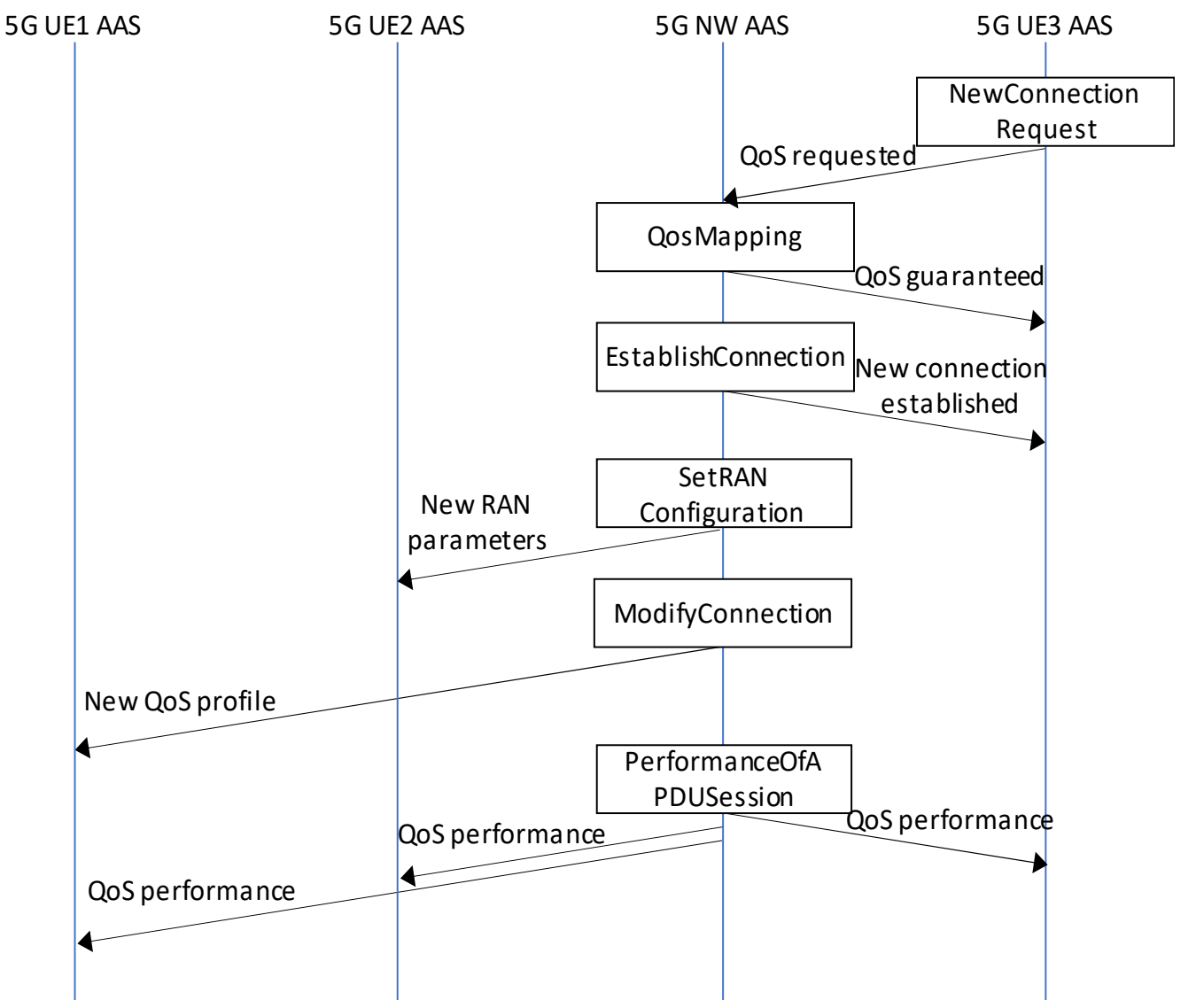

Fig. 4. QoS management use case using the 5G AAS.

## VII. Conclusions

This paper has presented the design and implementation of a 5G system AAS, including a 5G UE AAS and a 5G NW AAS. The design follows the 5G-ACIA guidelines to facilitate the integration of 5G with industrial systems and applications, as well as the Plattform Industrie 4.0 and 3GPP standards. We have adopted a functional design approach where information is structured and grouped by functions of a 5G network rather than by network nodes. This approach facilitates the scaling of the 5G AAS and its customization to the specific functionalities that are necessary or more relevant for each deployment of the 5G system AAS. The proposed 5G UE and NW AAS have been defined to provide and expose the data and capabilities of 5G necessary to support its integration with Industry 4.0 systems and applications, including, among others: UE identification, technical datasheets, configuration of virtual networks, connectivity parameters, location information, and QoS performance, monitoring and prediction capabilities. The defined AASs include passive properties and define active operations that can be tailored to future specific use cases where the AASs will be used to integrate 5G with manufacturing systems. The implementation of the 5G UE AAS and the 5G NW AAS presented in this paper is openly released to the community in [1].

## Acknowledgment

This work has been funded by European Union's Horizon Europe Research and Innovation programme under the Re4dy project (No 101058384), by MCIN/AEI/10.13039/ 501100011033 and the "European Union NextGenerationEU /PRTR" (TED2021-130436B-I00), by the Spanish Ministry of Science and Innovation and Universities, AEI and FEDER funds (EQC2018-004288-P), and by Generalitat Valenciana (CIGE/2022/17), and UMH's Vicerrectorado de Investigación grants (VIPROAS23/11 and 2024).

## References

[1] 5G-AAS, UWICORE lab., Universidad Miguel Hernandez de Elche (Spain). Github repository: https://github.com/uwicore/5G-AAS.

[2] *Programmable 5G for the Industrial IoT*, Ericsson Technology Review, Oct. 2022.

[3] Plattform Industrie 4.0, *Details of the Asset Administration Shell - Part 1*, Specification, v3.0RC02. May 30, 2022

[4] Q. Zhou, Y. Wu, S. He, Z. Shi and J. Chen, "A Unified Asset Management Architecture Based on OPC UA", in *Proc. of the China Automation Congress (CAC)*, Beijing, China, 2021, pp. 7605-7610.

[5] 5G-ACIA, *Using Digital Twins to Integrate 5G into Production Networks*, 5G-ACIA White Paper, Feb. 2021.

[6] G. Cainelli, L. Rauchhaupt, L. Underberg, "Supporting resilience with industrial 5G systems and Industrie 4.0", in *Proc. Jahreskolloquium "Kommunikation in der Automation* (KommA 2021), Magdeburg, Nov. 2021.

[7] G.P. Cainelli, L. Underberg, L. Rauchhaupt, C. E. Pereira, "Asset administration shell submodel for wireless communication system", *IFAC-PapersOnLine*, vol. 55, no 2, 2022, pp. 120-125.

[8] R. F. Ustok, A. Cihat Baktir and E. D. Biyar, "Asset Administration Shell as an Enabler of Intent-Based Networks for Industry 4.0 Automation", in *Proc. 2022 IEEE 27th International Conference on Emerging Technologies and Factory Automation* (ETFA), Stuttgart, Germany, 2022, pp. 1- 8.

[9] D. Rossi, G. Tontini, D. Borsatti and F. Callegati, "Integration of 5G connectivity with the Asset Administration Shell in Industry 4.0", in *Proc. of the 13th International Conference on Network of the Future (NoF)*, Ghent, Belgium, 2022, pp. 1-3.

[10] 5G-ACIA, *Exposure of 5G Capabilities for Connected Industries and Automation Applications*, 5G-ACIA White Paper, v2, Feb. 2021.

[11] AASX Package Explorer. Github repository: https://github.com/admin-shell-io/aasxpackageexplorer.

[12] Eclipse-basyx. Basyx-python-sdk. Github repository: https://github.com/eclipsebasyx/basyx-python-sdk.

[13] 3GPP, TSG SA; Procedures for the 5G System (5GS); Stage 2; Rel.17, 3GPP TS 23.502 V17.8.0, Mar. 2023.

[14] 3GPP, TSG CNT; 5G System; Network Data Analytics Services; Stage 3; Rel 17, 3GPP TS 29.520 V18.5.1; Apr. 2024.

[15] 3GPP, TSG SA; Service Enabler Architecture Layer for Verticals (SEAL); Functional architecture and information flows; Rel.17, 3GPP TS 23.434 V17.7.0, Sep. 2022.

[16] Plattform Industrie 4.0, *Details of the Asset Administration Shell, Part II*, Specification, v1.0RC02, Nov. 2021.

[17] Plattform Industrie 4.0, *AAS Reference Modelling*, Discussion paper, Apr. 2021.

[18] AASX Browser based on specifications of Platform Industrie 4.0, https://admin-shell-io.com/5011/.

[19] 3GPP, TSG SA; System architecture for the 5G System (5GS); Stage 2; Rel.17, 3GPP TS 23.501 V17.7.0, Dec. 2022.

[20] Generic Network Slice Template. GSM Association. Generic Network Slice Template Version 8.0 27 January 2023.

[21] 3GPP, TSG CNT; 5G System; Policy and Charging Control signalling flows and QoS parameter mapping; Stage 3; Rel.17, 3GPP TS 29.513 V17.10.0, Mar. 2023.

[22] 3GPP, TSGS SA; Service requirements for the 5G system; Stage 1; Rel 19, 3GPP TS 22.261, V19.6.0, April. 2024.

[23] 3GPP, TSG SA; 5G system (5GS) Location Services (LCS); Stage 2; Rel.17, 3GPP TS 23.273 V17.8.0, Mar. 2023.

[24] 3GPP, TSG RAN; NR; User Equipment (UE) radio transmission and reception; Part 1: Range 1 Standalone, Rel.17, 3GPP TS 38.101, V17.8.0, Dec. 2022.

[25] 5G-ACIA, *5G QoS for Industrial Automation*, 5G-ACIA White Paper, Nov. 2021.